# Feature Joint-State Posterior Estimation in Factorial Speech Processing Models using Deep Neural Networks

Mahdi Khademian, Mohammad Mehdi Homayounpour

**Abstract** This paper proposes a new method for calculating joint-state posteriors of mixed-audio features using deep neural networks to be used in factorial speech processing models. The joint-state posterior information is required in factorial models to perform joint-decoding. The novelty of this work is its architecture which enables the network to infer joint-state posteriors from the pairs of state posteriors of stereo features. This paper defines an objective function to solve an underdetermined system of equations, which is used by the network for extracting joint-state posteriors. It develops the required expressions for fine-tuning the network in a unified way. The experiments compare the proposed network decoding results to those of the vector Taylor series method and show 2.3% absolute performance improvement in the monaural speech separation and recognition challenge. This achievement is substantial when we consider the simplicity of joint-state posterior extraction provided by deep neural networks.

**Keywords** factorial speech processing models, deep neural networks, factorial hidden Markov models, state-conditional observation distribution, model combination using vector Taylor series, feature joint-state posterior

## 1. Introduction

Factorial speech processing models are powerful generative tools for modeling dynamics of two or more audio sources and the way they combine to generate mixed-audio signals. Fig. 1 shows the graphical model of factorial speech processing models in which two audio sources create a mixed-audio signal. In fact, these models are constructed based on factorial hidden Markov models (FHMM) [1], which consist of multiple Markov chains for modeling dynamic systems with multiple underlying independent stochastic processes.

Factorial speech processing models have been applied in the past for two main purposes: robust automatic speech recognition (ASR) in environments with non-stationary noises [2] and monaural multi-talker speech separation and/or recognition [3]. In the first scenario, two stochastic processes are speech process and non-stationary noise process, which are combine to produce the corrupted speech signal. For this case, factorial speech processing models are suggested for handling non-stationary noise situations [4] in which they provide improvement over their single noise-state robust-ASR techniques like model compensation [2,5]. In the second scenario, two or more speaker voices are mixed together to produce a multi-talker speech utterance in which underlying source processes are the speakers' voices. In this case, previous achievements are surprising [6,7], even better than the results achieved manually by human listening (Fig. 8-left). The main reason for this is the ability of the factorial models for generative modeling of the mixing procedure in which the models are best

Mahdi Khademian, Mohammad Mehdi Homayounpour*
Laboratory for Intelligent Multimedia Processing (LIMP), Amirkabir University of Technology, Tehran, Islamic Republic of Iran
email: homayoun@aut.ac.ir, Tel:+982164542722, Fax: +982164542700

fitted to this application. Information flow and exchange in source speech Markov chains is very effective for resolving ambiguities during speech decoding since only one recording channel is available in this application.

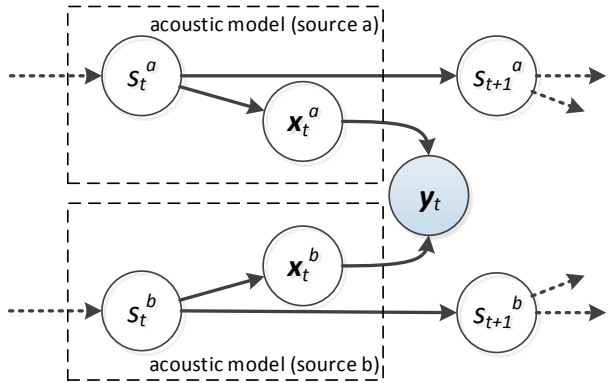

**Fig. 1. Graphical model of factorial speech processing models. This figure shows all variables of frame $t$ and only shows state variables of frame $t + 1$. It shows two audio sources, which are related by $p(y_t|x_t^a, x_t^b)$ to generate the mixed-audio feature of frame $t$. Source acoustic models are simply independent hidden Markov models for each audio source.**

The information flow requires having joint-state likelihoods (or posteriors) for the mixed-audio features, i.e., $p(y_t|s_t^a, s_t^b)$ or $p(s_t^a, s_t^b|y_t)$. There exist at least two problems with providing joint-state likelihoods for the decoder. First, the calculation of joint-state likelihoods requires a potentially squared number of states for which the likelihood should be calculated. This significantly increases the required calculations during the decoding. Second, there are some approximations involved in most of the methods for calculating joint-state likelihood. For example, in the max model, the max approximation is valid for high resolution spectral features [6], while for recognition purposes, we need more elaborated feature spaces [4]. As another example, the data-driven parallel model combination (PMC) and vector Taylor series (VTS) methods rely on mismatch functions where there is at least one approximation involved in developing the mismatch functions [8,9]. The VTS method also uses a linear approximation of the nonlinear mismatch function. Considering the third problem, the data-driven PMC and the method of weighted stereo samples (WSS) [5] need an additional parametric modeling step along with saving joint-state parametric models, which require quadratic space compared to the separate hidden Markov model (HMM) acoustic models. Moreover, while the WSS method provides the most accurate results since it does not rely on the mismatch function, it suffers from a need for stereo data.

Recent efforts make the deep neural networks (DNN) a powerful acoustic model in ASR applications, which is referred to as DNN/HMM architecture [10,11]. In these works, DNNs are used to extract senone posteriors. The senone posteriors are then passed to the decoder that performs the decoding. In this architecture, hidden Markov models (HMM) are only used for decoding and not extracting the acoustic likelihoods. A recent work by Microsoft research [12] applies DNNs to a task that is closely related to factorial models: the monaural speech separation and recognition challenge [13]. In this challenge, we consider two chains in the factorial model in which each models a speaker, while both speakers simultaneously issue a simple command. In [12], two separate DNNs are used for extracting marginal posteriors that are used in a decoder. In this case, there cannot be any information flow between the two chains of the factorial model. The state posteriors must be extracted in the joint form to make the information flow possible between chains of factorial models.

This paper proposes a new method for extracting true joint-state posteriors to be used in factorial speech processing models. This way of applying DNNs along with using factorial speech processing models is comparable to applying DNNs in the DNN/HMMs [11]. Therefore, by proposing the mentioned method, this work brings the DNNs to be used in factorial models as a DNN/FHMM architecture. This architecture can be used in robust-ASR applications that require further development of the proposed method.

The rest of this paper is organized as follows. The next section describes the exact method for the calculation of joint-state likelihoods and a variety of approximated estimation and modeling techniques for this purpose. It simplifies presenting the current methods in a unified view to bring a clear insight into the problem. The third section describes the proposed method for calculating joint-state posteriors using deep neural networks. It defines the objective function, the way to optimize it, and some practical considerations for training the network. Section 4 describes the experiment framework, dataset, and achieved results, and the last section concludes the paper.

## 2. Methods for calculating features' joint-state likelihoods

The role of joint-state likelihoods in factorial models of speech processing is like the observation distribution of HMMs, which is required for decoding. The following expression exactly calculates the joint-state likelihood of $\mathbf{y}_t$ (the time index, $t$ is omitted for simplicity here and where it is not important to be mentioned):

$$p(\mathbf{y}|s^a, s^b) = \iint p(\mathbf{y}|\mathbf{x}^a, \mathbf{x}^b) p(\mathbf{x}^a|s^a) p(\mathbf{x}^b|s^b) d\mathbf{x}^a d\mathbf{x}^b \tag{1}$$

where $p(\mathbf{x}^a|s^a)$ and $p(\mathbf{x}^b|s^b)$ are the source state conditional likelihoods of the two audio sources. After this calculation, audio source features ($\mathbf{x}^a$ and $\mathbf{x}^b$) are marginalized out and removed from the graphical model of Fig. 1. This intermediate step is required for performing joint-decoding over the factorial models for speech recognition.

The exact calculation of (1) is not easily possible and is highly related to the feature space. Actually, in the exact calculation of (1), we faced a dilemma: having more accurate acoustic interaction models, i.e., $p(\mathbf{y}|\mathbf{x}^a, \mathbf{x}^b)$, make direct calculation of (1) impossible; using an approximate interaction model leads to an inaccurate joint-state likelihood. In the first case, we must approximate the marginalization and in the second case even with exact marginalization, the calculated joint-state likelihood is based on the approximated interaction model. Several approaches were developed in the past for a direct and implicit calculation of (1), which are mainly used in robust-ASR applications [4,5]. We briefly describe four main approaches for calculation or estimation of joint-state likelihoods of acoustic features with different characteristics. In these approaches, the following time domain expression is considered to mix the two audio source signals:

$$y = x^a + x^b \tag{2}$$

This expression can be elaborated to support robustness against the audio sources' relative energy change and the environment's channel response [8]; here, we consider it in its simplest form, since this brings the focus on the main characteristics of provided approaches for joint-state likelihood calculation.

### 2.1. Joint-state likelihood calculation using the max-model

For high-resolution features like the log-power spectrum, the following interaction function is an approximation for the sum of the two source audio signals (expression (2)):

$$p(\mathbf{y}|\mathbf{x}^a, \mathbf{x}^b) = \boldsymbol{\delta}_{\mathbf{y}-\max(\mathbf{x}^a, \mathbf{x}^b)} \tag{3}$$

in which $\boldsymbol{\delta}_{(.)}$ is a Dirac delta function and maximization is performed element-wise for each of two frequency bins of the source feature vectors. Using a single Gaussian for modeling each conditional source feature vector, the following expression is derived based on (3) [6]. This joint-state likelihood is calculated exactly, although it is based on the early approximation of (3) for the acoustic interaction model.

$$p(\boldsymbol{y}|s^a, s^b) = p_{x^a}(\boldsymbol{y}|s^a)\boldsymbol{\Phi}_{x^b}(\boldsymbol{y}|s^b) + p_{x^b}(\boldsymbol{y}|s^b)\boldsymbol{\Phi}_{x^a}(\boldsymbol{y}|s^a) \qquad (4)$$

In (4), $x^a$ and $x^b$ subscripts denote the stochastic variable and $\boldsymbol{\Phi}$ is the Gaussian cumulative distribution function (CDF) of the denoted variable. Due to the required calculations for numerical methods for evaluating CDFs of Gaussians with full covariance matrices, in practice, joint-state likelihoods are calculated for audio source models with diagonal covariance matrices [6]. This simplification reduces the accuracy of audio source models, especially for high resolution spectral features.

## 2.2. Joint-state conditional distribution modeling using state conditional samples (known as data-driven parallel model combination)

The parallel model combination, PMC, is a model-based noise-robust speech processing method in which corrupted speech distribution is assumed to be Gaussian in the log spectrum domain. Based on this assumption, the parameters of clean source models are adjusted for providing robustness against environmental changes during the test.

In the data-driven PMC methods, the normality assumption exists, but the parameters of the normal distribution are estimated by a Monte Carlo method. In these methods, state-conditional samples of audio source features are evaluated in a function, called "mismatch function", to generate corrupted-speech features. The mean vector and covariance matrix of the state conditional distribution are then estimated based on the stated-conditional corrupted samples. Here, we briefly derive the mismatch function for the Mel-Cepstral domain, which will be used in our experiments.

Using the short-term discrete time Fourier transform, the following relationship exists between each frame of the audio source feature vectors and the mixed-signal feature in the power spectrum when the two sources are additively combined using the expression (2):

$$|Y|^2 = |X^a|^2 + |X^b|^2 + 2|X^a| \circ |X^b| \circ \cos \boldsymbol{\theta} \qquad (5)$$

where $\boldsymbol{\theta}$ is the vector of phase difference across different frequency bins of the source features. The above relationship is the simplest mismatch function that can be used in noise-robust ASR methods when considering one audio source as speech and the other one as noise. For the Mel-Cepstral domain, we first extract the feature using two linear and a non-linear operation as follows:

$$\boldsymbol{y} = \mathbf{C} \log(\mathbf{W}|Y|^2) \qquad (6)$$

in which $\boldsymbol{y}$ are Mel-Frequency Cepstral Coefficients (MFCC) extracted from the high-resolution complex spectrum. Moreover, $\mathbf{C}$ is the discrete cosine transform (DCT) matrix and $\mathbf{W}$ contains Mel-scale averaging filters. Using the pseudo-inverse DCT matrix and exponential operator, we can retrieve Mel-power spectrum coefficients from MFCC features. In this domain, the following relationship is held between the feature vectors:

$$|\bar{Y}|^2 = |\bar{X}^a|^2 + |\bar{X}^b|^2 + 2\boldsymbol{\alpha} \circ |\bar{X}^a| \circ |\bar{X}^b| \qquad (7)$$

where the feature vectors are extracted simply by Mel-scale averaging of power spectral features, i.e., $|\bar{Y}|^2 = \mathbf{W}|Y|^2$. In addition, $\boldsymbol{\alpha}$ is a stochastic vector over the interval of $[-1, 1]$ and is called phase factor, which reflects the effect of phase difference between complex discrete Fourier vectors ($\boldsymbol{\theta}$ in (5)). The phase factor is usually considered as a constant and may even be considered to be zero in some applications [9]. We consider it zero in the rest of this paper. Using one forward and two inverse MFCC operators (consecutive use of logarithm and DCT matrix as the forward operator), the following mismatch function is extracted for the MFCC static features:

$$\boldsymbol{y} = \mathbf{C}\log\bigl(\exp(\mathbf{C}^{-1}\boldsymbol{x}^a) + \exp(\mathbf{C}^{-1}\boldsymbol{x}^b)\bigr) \qquad (8)$$

in which $\mathbf{C}^{-1}$ is the pseudo inverse of the DCT matrix. For detailed derivation and generalizations, the reader can refer to [8,9]. Using mismatch functions like (8), we can extract mixed-audio feature samples from the source-audio features. These samples can be used in parametric modeling of the joint-state observation distribution of models in the form of Fig. 1.

## 2.3. Joint-state likelihood estimation using vector Taylor series

In the method of vector Taylor series for evaluating state likelihoods, the joint-state distribution is interpreted as the linear transformation of source variables. For explanation, consider we have two sets of audio source models. Assume that we have two sets of HMMs for audio sources in models in the form of Fig. 1. Now, each state conditional observation distribution in HMMs is one audio source stochastic variable, i.e., $\boldsymbol{x}^a$ or $\boldsymbol{x}^b$. Using the linear approximation of (8), these two variables can be transformed into the joint-state conditional observation distribution in the factorial model. The first order vector Taylor series is used for linearization of (8) at $\langle \boldsymbol{x}_0^a, \boldsymbol{x}_0^b \rangle$ as follows:

$$\boldsymbol{y} \cong \mathbf{C}\log\bigl(\exp(\mathbf{C}^{-1}\boldsymbol{x}_0^a) + \exp(\mathbf{C}^{-1}\boldsymbol{x}_0^b)\bigr) + \mathbf{J}_{\boldsymbol{y}_{x^a}}(\boldsymbol{x}^a - \boldsymbol{x}_0^a) + \mathbf{J}_{\boldsymbol{y}_{x^b}}(\boldsymbol{x}^b - \boldsymbol{x}_0^b) \qquad (9)$$

in which $\mathbf{J}_{\boldsymbol{y}_{x^a}}$ and $\mathbf{J}_{\boldsymbol{y}_{x^a}}$ are Jacobian matrices of the mismatch function relative to $\boldsymbol{x}^a$ and $\boldsymbol{x}^b$ evaluated at the expansion points. The expansion points are usually selected as the means of source variables.

At this step, we can see (9) as a linear transformation of two multivariate Gaussians, which yields to another Gaussian variable, $\boldsymbol{y}$. The mean and covariance matrix of $\boldsymbol{y}$ can be extracted by a linear combination of means and covariance matrices of $\boldsymbol{x}^a$ and $\boldsymbol{x}^b$. Returning to the original problem, if we have two state-conditional audio source variables, we can have an approximation of joint-state conditional observation distribution, which will be used in decoding.

## 2.4. Joint-state distribution modeling using weighted stereo samples

The method for joint-state distribution modeling using weighted stereo samples (WSS method), is a Monte Carlo method for modeling joint-state conditional observation distribution [5]. In this sense, the method is like data-driven PMC methods, but it does not rely on any mismatch function. Instead, it uses weighted stereo samples to model the observation distribution in factorial models, parametrically. The following expression constructs an empirical distribution for the joint-state conditional observation distribution in the WSS method:

$$p(\boldsymbol{y}|s^a, s^b) = \sum_{k=1}^{N} w_{k|s^a} w_{k|s^b} \boldsymbol{\delta}(\boldsymbol{y}_k - \boldsymbol{y}) \qquad (10)$$

in which $\boldsymbol{y}_k$ are the mixed-speech features that are paired with audio source features and their source-state weights are $w_{k|s^a}$ and $w_{k|s^b}$. The weights are calculated by the following relationships using the paired audio source features ($\boldsymbol{y}_k \leftrightarrow \boldsymbol{x}_k^a$ and $\boldsymbol{y}_k \leftrightarrow \boldsymbol{x}_k^b$):

$$w_{k|s^a} = p(x_k^a|s^a)/\sum_{s^a} p(x_k^a|s^a)p(s^a) \qquad (11)$$

$$w_{k|s^b} = p(x_k^b|s^b)/\sum_{s^b} p(x_k^b|s^b)p(s^b) \qquad (12)$$

There is also a developed method for modeling the observation distribution parametrically using the weighted samples [5]. Using the trained parametric model, we can calculate joint-state likelihoods for decoding.

## 3. Joint-state posterior estimation using deep neural networks

The effectiveness of factorial speech processing models in the speech processing tasks relies on the ability to have information flow between its multiple chains and use of this information flow during the inference. This requirement is satisfied by providing joint-state likelihoods or posteriors of the mixed-audio features. The requirement exists here since there is a V-shape connection between audio-sources and mixed features in the Fig. 1 graphical model. As mentioned earlier, even with the development of various methods for calculation of joint-state likelihoods, there remains an inability to perform fast and accurate evaluation of these likelihoods during the inference.

The objective of the proposed network is to extract joint-state posteriors of the mixed-audio features, $p(s^a, s^b|\boldsymbol{y}_t)$, which can be used in the decoder. Naively, this can be done by training a deep feed forward network to extract joint-state posteriors using training samples calculated using existing joint-state likelihood estimation techniques. It means that in the output layer, the network must have $s^a \times s^b$ neurons to generate the posteriors. By training such a network and using it along with a joint decoder, we can apply a DNN/FHMM architecture to be used for robust and multi-talker speech recognition applications.

The problem for training the proposed network is that the joint-state likelihoods are not available due to the difficulties involved in the calculation of (1). Even by accepting the significant required calculations involved in the current methods for providing the network training data, the approximations involved in data preparation may not be acceptable.

The solution is to use marginal posteriors of the mixed-audio features, then estimate joint-posteriors from the marginal posteriors. At first, this method seems impossible since it requires the need to solve the following underdetermined system of equations:

$$\begin{aligned} \mathbf{X} \cdot \mathbf{1}_{|s^b| \times 1} &= \boldsymbol{p}(s^a|\boldsymbol{y}) = \boldsymbol{m}^a \\ \mathbf{X}^\mathrm{T} \cdot \mathbf{1}_{|s^a| \times 1} &= \boldsymbol{p}(s^b|\boldsymbol{y}) = \boldsymbol{m}^b \end{aligned} \qquad (13)$$

in which $\boldsymbol{p}(s^a|\boldsymbol{y})$ is the vector of state marginal posteriors conditioned on $\boldsymbol{y}$ for the audio source $a$, and $\boldsymbol{p}(s^b|\boldsymbol{y})$ is for the audio source $b$ and $\mathbf{X} \triangleq [x_{i,j} = p(s^a = i, s^b = j|\boldsymbol{y})]$. In (13), two vectors of ones perform marginalization over rows and columns of $\mathbf{X}$.

The problem for solving the underdetermined system of equations of (13) can be solved by the following initialization. At the beginning of the network fine-tuning phase, we can temporarily use approximate joint-posteriors to bring the network to a feasible solution for the equations. The discriminative training phase can then be followed by altering joint-posteriors to generate marginal posteriors. In fact, we assume that the initial solution to the system of equations with approximate joint-posteriors can be improved iteratively during the discriminative phase using marginal posteriors. Based on this assumption, we propose the following three steps for training a deep neural network for extracting joint-state posteriors: the generative phase, initializing joint-state layer weights, and fine-tuning the network. Fig. 2 provides the network architecture illustrating these steps. These steps are described in the following sub-sections.

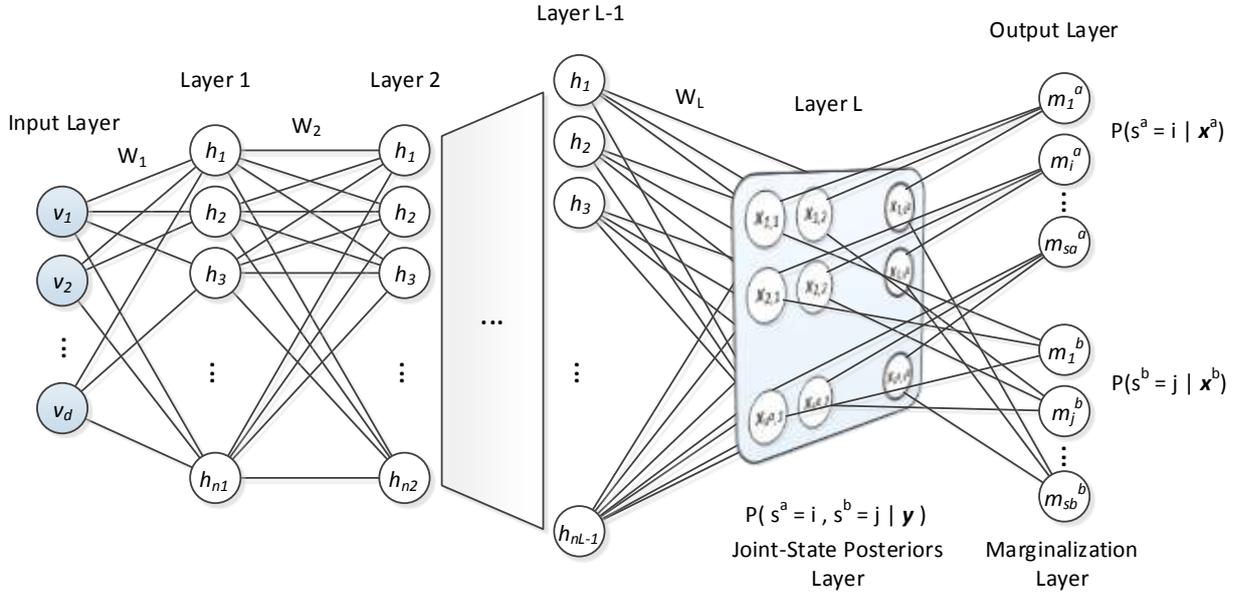

Fig. 2. Deep architecture for extracting the joint-state posteriors. From layer 1 to layer L-1, the network is pre-trained layer-by-layer generatively in the first phase. In the second phase, weights of layer L are initialized for generating approximate joint-state posteriors. In the third phase, weights of layer 1 to layer L are fine-tuned so that the network can generate marginal posteriors close to the pairs of state posteriors.

### 3.1. The generative phase

In the first step, unlabeled mixed-audio features will be used to train a stack of restricted Boltzmann machines (RBM), which is used to create a deep belief network (DBN), thus performing initialization of the first L-1 layers of the network in Fig. 2. The following data log-likelihood is maximized in this step for each RBM using the persistent contrastive divergence (PCD) method [14].

$$\sum_{k=1}^{N} \log p(v_{k,l}) = \sum_{k=1}^{N} \log \frac{1}{Z} \sum_{h} e^{-(-a_l^T v_{k,l} - b_l^T h_{k,l} - v_{k,l}^T \widehat{W}_l h_{k,l})} \tag{14}$$

In (14), $N$ is the number of training data, $v_{i,l}$ are visible data for the $l$th layer, $h_{k,l}$ are layer $l$'s extracted hidden samples from its visible data, and $a_l$, $b_l$ and $\widehat{W}_l$ are parameters of the $l$th RBM. In (14), the inner summation is done for all combinations of the hidden neurons, but in the PCD method, there is no explicit calculation of this summation. Instead, a series of hidden samples is extracted from the visible data, which is used for gradient calculation. The approximated gradients are used for updating the parameters for optimizing the objective function of (14).

Like successful use of DNNs in speech recognition applications, we use a context window of raw speech features to feed the first RBM as its visible data (see Fig. 3). Extracted features of the first RBM provide the second RBM training data, and this procedure is followed layer by layer until the final RBM. At each step, the input data's likelihood increases as the model extracts higher level features from the raw data [14,15].

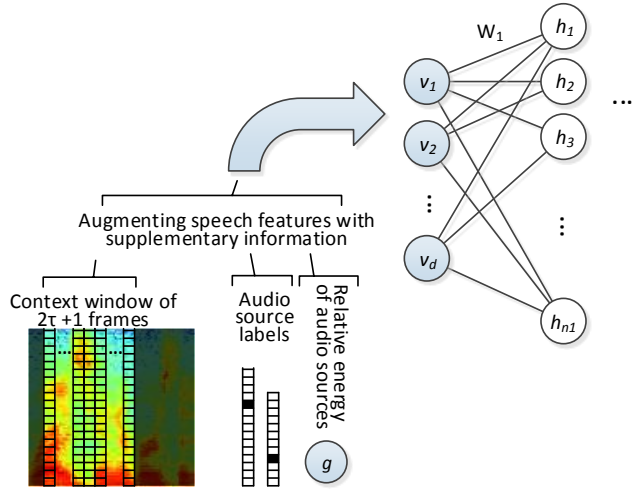

Fig. 3. Composition of the network input data.

The context window of raw features can be augmented with supplementary information, which further improves the DNN classification. In the tasks where the audio sources are the same type, any discrimination provided by this supplementary information can reduce joint-state swapping (transposition of the joint-state posterior matrix). For example, consider joint-speech recognition in a task in which both audio sources are two speakers selected from a close set of speakers. In such a task, we can augment the raw features with joint-speaker identification data.

### 3.2. Initialization of the joint-state layer weights

The objective of the second training phase is to provide a feasible solution to the system of equations constructed by the output layer (marginalization layer in Fig. 2). This is done by training weights of layer L (also tuning weights of layer 1 through L–1) for generating approximate joint-state posteriors.

At first, the DBN is converted to a DNN by using hidden unit biases ($b_l$) and RBM weights (visible unit biases are discarded). Therefore, the constructed DNN consists of layers of pre-trained weights as $\mathbf{W}_l = \begin{bmatrix} \widehat{\mathbf{W}}_l \\ \boldsymbol{b}_l \end{bmatrix}$ connecting to a series of sigmoid neurons. The neurons of layer L are selected to have sigmoid activation function where the range of sigmoid function supports the range of posterior probabilities. Training the network up to layer L is done by the stochastic gradient decent algorithm, which minimizes the following objective function:

$$\frac{1}{2}\sum_{k=1}^{N}\sum_{j=1}^{|s^b|}\sum_{i=1}^{|s^a|}\left(\mathbf{X}_k(i,j) - \mathbf{D}_k(i,j)\right)^2 \tag{15}$$

in which $N$ is the number of training data, $\mathbf{X}_k$ are network outputs (from the layer L, see Fig. 2), and $\mathbf{D}_k$ are data labels (training is done over mini-batches). The data labels for this training phase can be provided using approximate joint-state posteriors extracted by an approximated method. At first, the selected method calculates joint-state likelihoods as follows:

$$p(\boldsymbol{y}_k|s^a = i, s^b = j) = \mathcal{F}(\boldsymbol{y}_k, \mu_i, \Sigma_i, \mu_j, \Sigma_j) \tag{16}$$

in which, $\mathcal{F}$ is the selected method for approximating the joint-state likelihoods. In our experiments, we use the VTS method for preparing these labels as presented in [7]. For example, for observation distribution of HMMs, joint-state conditional parameters of source models are $\mu_i, \Sigma_i, \mu_j, \Sigma_j$. These parameters are first combined using the VTS method for each joint-state. The likelihood of mixed-speech feature vector ($\boldsymbol{y}_k$) is then evaluated for each joint-state. Joint-state posteriors can then be calculated after the evaluation of all joint-state likelihoods assuming all states are equally likely. Therefore, we have data labels as follows:

$$\mathbf{D}_k(i,j) \propto p(\mathbf{y}_k|s^a = i, s^b = j) = \mathcal{F}(\mathbf{y}_k, \mu_i, \Sigma_i, \mu_j, \Sigma_j) \tag{17}$$

The matrices of data labels are used for initializing the DNN up to layer L with the approximated joint-state posteriors.

### 3.3. Fine-tuning the network

The main training phase of the network is to use marginal posteriors provided at the output layer for tuning network weights to extract joint-state posteriors at layer L. The extracted joint-state posteriors are used for joint-decoding. Since we do not have marginal posteriors of mixed-audio features, we accept an approximation to use state posteriors of stereo audio features of the sources paired to the mixed-audio features. The exact state posteriors of the sources are calculated for the stereo audio features using the following expressions:

$$d\mathbf{m}^a_{s^a} \triangleq p(s^a|\mathbf{x}^a) = p(\mathbf{x}^a|s^a)p(s^a)/\sum_{s^a} p(\mathbf{x}^a|s^a)p(s^a) \tag{18}$$

$$d\mathbf{m}^b_{s^b} \triangleq p(s^b|\mathbf{x}^b) = p(\mathbf{x}^b|s^b)p(s^b)/\sum_{s^b} p(\mathbf{x}^b|s^b)p(s^b) \tag{19}$$

In fact, source audio features are evaluated at their corresponding source models for extracting the state conditional likelihoods. Assuming all source states have the same prior, the state posteriors are proportional to the likelihoods, i.e., $p(s^a|\mathbf{x}^a) \propto p(\mathbf{x}^a|s^a)$.

In a unified view, the pairs of state posteriors provided by (18) and (19) are the desired values for training the whole network, and the following objective function is minimized for this training phase:

$$J = \frac{1}{2}\sum_{k=1}^{N} \left( \sum_{i=1}^{|s^a|} \left( \sum_{j=1}^{|s^b|} \mathbf{X}_k(i,j) - p(s^a = i|\mathbf{x}^a) \right)^2 + \sum_{j=1}^{|s^b|} \left( \sum_{i=1}^{|s^a|} \mathbf{X}_k(i,j) - p(s^b = j|\mathbf{x}^b) \right)^2 \right) \tag{20}$$

in a vector form, we have:

$$J = \frac{1}{2}\sum_{k=1}^{N} \left( \|\mathbf{m}^a_k - d\mathbf{m}^a_k\|^2 + \|\mathbf{m}^b_k - d\mathbf{m}^b_k\|^2 \right) \tag{21}$$

where $\mathbf{m}^a_k$ and $\mathbf{m}^b_k$ are the extracted marginalized posteriors, e.g., $\mathbf{m}^a_k(i) = \sum_{j=1}^{|s^b|} \mathbf{X}_k(i,j)$. The partial derivatives of the objective function with respect to network outputs at layer L are as follows:

$$\frac{\partial J}{\partial \mathbf{X}(i,j)} = \left( \mathbf{m}^a(i) - d\mathbf{m}^a(i) \right) + \left( \mathbf{m}^b(j) - d\mathbf{m}^b(j) \right) \tag{22}$$

These partial derivatives are propagated to update network weights from layer L to layer 1 using the chain rule.

## 4. Experiments

The proposed method of this work for extracting joint-state posteriors is evaluated using the dataset of the "monaural speech separation and recognition challenge" [13]. The main task of this dataset is to recognize the spoken words of a designated speaker from a mixed-speech signal of two speakers. More specifically, the task is defined based on the grammar illustrated in Fig. 4. In this challenge, two speakers are simultaneously issuing a simple command extracted from a pre-defined grammar.

One of the two speakers, called the target speaker, always uses the word "white" as his mentioned color in the uttered command and the other does not. In the challenge, the standard score is calculated based on the recognized "letter" and "number" of the target speaker. The final recognition accuracy of this task is calculated only using these two-word slots as provided by the task standard scoring scripts.

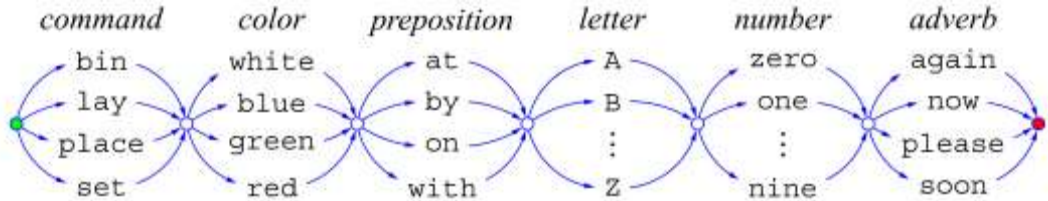

**Fig. 4. The grammar of "monaural speech separation and recognition challenge" [6]. The target speaker always uses "white" as his mentioned color and the masker does not.**

The dataset is created based on single speaker commands of 34 speakers in which each speaker issues 1,000 different commands generated based on the mentioned grammar. From each speaker, 500 utterances are selected and used as the training set. The other 500 utterances are used to create the test set as follows. Two utterances are randomly selected from this set in which one utterance contains the word "white" and the other does not. These two utterances are combined using the following time-domain relation:

$$y[t] = x^a[t] + gx^b[t] \qquad (23)$$

where $x^a$ and $x^b$ are the utterances of the target and masker speakers and $g$ is the masker speaker gain.

The masker speaker gain is adjusted for creating multiple instances of the mixed-speech signal with different relative energy of the target and masker speech. This is called target to masker ratio (TMR) and is expressed in dB scale. From each of the selected pair of utterances, six different mixed-speech signals are created with TMRs of 6, 3, 0, -3, -6, and -9 dBs. It is expected that recognizing target speaker command becomes more difficult in low TMRs, which is apparent for the single-talker baseline (see single-talker recognition performance in Fig. 6). The test set contains 600 mixed-speech signals for each TMR and a set of 600 utterances (called the clean set), which consist of only the target utterance (TMR $= \infty$).

### 4.1. Training source models and the single-talker recognition system

In this work, one-state monophones are used as acoustic models. The decoding is done using these models, task grammar (Fig. 4), and lexicon of the task words[1]. Mel-Frequency Cepstral coefficients are used for source model features with 27 filters in the Mel-scale filter bank, and the first 19 Cepstral coefficients (including the 0th coefficient) are used as the static feature. The static features are augmented with their first order derivative. Therefore, a total of 38 coefficients are used for source modeling. Finally, the source models are adapted for each of the 34 speakers, which is used for speaker dependent label extraction for training the network.

The standard task score for the clean folder of the test set is recognized by the single-talker recognition system across different source models and is provided in Table 1. In fact, the only difference between source models of this table is the number of Gaussian components for acoustic modeling of each phoneme using Gaussian Mixture Models (GMM).

---
[1] - The task contains 51 words which its transcription is extracted from the BEEP dictionary.

Table 1: Recognition results for the single-talker recognition system over the clean set of the challenge test set.

| Number of Source Model Components | Same Talker | Same Gender | Different Gender | Average |
|---|---|---|---|---|
| Single Gaussian | 66.29 % | 70.39 % | 67.75 % | 68.00 % |
| 4 Component GMMs | 83.26 % | 84.36 % | 82.25 % | 83.25 % |
| 8 Component GMMs | 89.82 % | 88.27 % | 88.00 % | 88.75 % |
| 64 Component GMMs | 90.50 % | 90.78 % | 92.00 % | 91.08 % |

Recognition results of the task for the 64-component single-talker recognition system are provided in Fig. 6. Even the best single-talker system performance is far from the factorial model results using the VTS method for calculation of joint-state likelihoods (Fig. 6). As another observation, we see that the recognition performance is degraded in low TMR conditions in which the target speaker voice is highly masked by the masker speaker.

### 4.2. Joint-decoding

The joint-decoder objective is to find the most probable word sequences of the two speakers given features of the mixed-audio signal; we use all allowed information during the decoding in which the target speaker uses the word "white" in his command and the masker does not. The decoder solves the following optimization problem:

$$w_{1:6}^{a*}, w_{1:6}^{b*} = \underset{w_{1:6}^{a}, w_{1:6}^{b}}{\mathrm{argmax}}(w_{1:6}^{a}, w_{1:6}^{b} | \mathbf{y}_{1:T})$$

$$st:\ w_2^a = \{\text{white}\}, w_2^b = \{\text{blue} \mid \text{red} \mid \text{green}\}$$

(24)

Decoding of this task is done using a joint decoder implemented by the joint token passing algorithm proposed by [7]. The decoder reads HTK HParse generated wordnets generated based on the task grammar. Fig. 5 provides the grammar for the masker speaker.

Two wordnets are used by the decoder in conjunction with a dictionary to perform the simultaneous decoding of the target and masker utterances presented in the mixed-audio signal. The decoder only requires joint-state posteriors of frames of the mixed-audio signal for its input. The joint-state posteriors are stored in a three-dimensional array in which its first dimension is the frame index and the two others form the joint-state posterior matrices. In the experiments, joint-state posteriors are provided for the decoder by two methods: the VTS method and the proposed deep neural network. In fact, the combination of the proposed deep neural network for providing joint-state posteriors and a joint-decoder suggests a DNN/FHMM architecture for speech recognition applications. Here, we apply this architecture for multi-talker speech recognition.

The decoder output is the most probable word sequences of the two speakers, i.e., $w_{1:6}^{a*}$ and $w_{1:6}^{b*}$. Finally, the task standard score is calculated based on the recognized "letter" and "number" of the target speaker using the scripts provided in the challenge resources.

```
$command = bin | lay | place | set;
$color = blue | green | red;
$preposition = at | by | in | with;
$letter = a | b | c | d | e | f | g | h | i | j | k | l | m | n | o | p | q | r | s | t | u | v | x | y | z;
$number = one | two | three | four | five | six | seven | eight | nine | zero;
$adverb = again | now | please | soon;

(sil $command $color $preposition $letter $number $adverb sil)
```

Fig. 5. The grammar defined using the HTK HParse tool for generating the wordnets [16]. This grammar is defined based on the task constraints (provided in Fig. 4) using HTK grammar definition language. The decoder uses the generated wordnet of this grammar for the masker speaker. For the target speaker, the color word-slot only contains the word "white".

### 4.3. The VTS baseline system

In the VTS baseline system, joint-state likelihoods are calculated by the VTS method and are used for the joint-decoding similar to implementation of [7]. Three-dimensional arrays of joint-state likelihoods are passed into the decoder for decoding. First, two joint-state conditional source models are combined using the VTS method, then feature likelihoods are evaluated in each joint-state combined model. The model combination is done for static and dynamic parts of features as mentioned in [8].

The decoding is done for source models with only four-component GMMs due to the significant amount of calculations involved during the joint-state likelihood extraction. The joint-state likelihoods are calculated without considering the phase factor (see equations (5) and (7)), and the decoding is done with full beam size.

The standard task score using the VTS based joint-decoder is provided for the two test scenarios in Fig. 6 as the baseline system: recognition using the gain adapted masker models and without adapted masker models. Gain adaptation requires estimation of relative energy of the target and masker speeches. The gain estimation and model adaptation are done similar to [7]. The reason for adapting masker models is that without this adaptation, the recognition results degrade significantly in the extreme TMR conditions (6, -6, and -9 dB) as shown in Fig. 6. In this figure, we see that joint-decoding without gain adaptation yields to performance degradation in extreme TMR conditions because in these conditions, audio source models for the masker chain are not matched for the test utterances (all source models are trained in near zero TMR conditions).

In the main experiments, the DNNs are trained using multi-TMR conditions. This improves their robustness to change in different TMR conditions (this will be investigated in the experiments). Finally, for the baseline system, the speaker identification information is extracted from the test set filenames as in the main experiments. The speaker identification is not directly related to the objective of this paper and is investigated in other works [6,7].

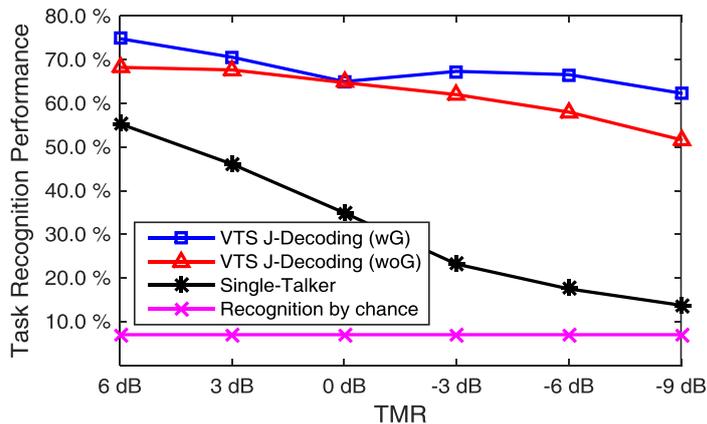

**Fig. 6. Baseline system recognition results.**

### 4.4. The DNN based system

The proposed network architecture is evaluated by training a DBN with the following hidden layer sizes in the generative phase: [2025, 2500, 3600, 5625]. In fact, these four hidden layers are layers 1 to L – 1 of the network in Fig. 2. Multiple experiments suggest increasing the number of hidden neurons just before the joint-state posterior layer. The DBN of our work is trained using the DeeBNet toolbox [15].

Network input is constructed using a context window of 17 frames of 50 Mel-scale filterbank augmented by speaker identification information. For generating mixed-speech utterances, two utterances are randomly selected from the clean speech data of 34 speakers and mixed according to (23). About 20,000 mixed-speech utterances are created from the training set of clean speech data and used for DNN training.

The mixed-audio signals are generated for two scenarios. In the first scenario, a DNN is trained for extracting joint-state posteriors without adjusting utterance gains. In the main scenario, a DNN is trained in a multi-style manner using gain adapted masker utterances for creating mixed-audio signals. Additionally, the gain information is added to the network input data, the effect of which will be discussed in the experiments.

For initializing joint-state layer weights, VTS approximated joint-state posteriors are calculated using (17). The speaker adapted models and the synthetized mixed-speech signals are used for this purpose. Half of the mixed-speech signals (10,000 files) are used for initialization of joint-state layer weights. Since we have 39 phonemes (one state for each) and a silence model in the acoustic model inventory, the total number of states for each chain equals 40. Thereafter, the number of joint-state neurons in the fifth layer of the network equals 1,600. This is the reason for keeping acoustic models as simple as possible to prevent the quadratic increase of neurons in the joint-state layer.

The network is fine-tuned using partial derivatives of the objective function (21) with respect to network parameters. These partial derivatives are calculated using the chain rule and by back-propagating error signals of expression (22). The error signals are calculated using the desired exact state posteriors and the current network extracted marginal posteriors. The exact state posteriors of the stereo features are calculated using (18) and (19). The feature likelihoods are evaluated in the two audio source models with 64-component GMMs. In this phase, the whole network is fine-tuned only using the two sets of state posteriors to generate the joint-state posteriors of the mixed-audio features. All the training files are used in this phase with their corresponding state posteriors. Initialization of the joint-state layer weights and fine-tuning of the network are done using a modification to the GPU version of the DeepLearnToolbox[2]. Fig. 7 provides recognition results of the joint-decoder for the DNN joint-state posterior extractor.

In "DNN J-Decoding", the network is trained using gain unaltered target and masker utterances. In the multi-gain DNN, training data is generated using gain adjusted masker utterances in a multi-style training fashion to better represent test conditions. Taking the masker gain into account is significant in the same talker conditions because in such conditions, the main discriminative feature of the two sources is the relative gain of the target and masker utterances. In other joint-speaker conditions, speaker and gender information is used to discriminate two sources, which improves joint-decoding results. When the gain is used as discriminative information (both in the DNN and VTS scenarios), the decoding performance increases, even in low TMR conditions.

For the gain aware cases, the main observation from Fig. 7 is that the DNN exploits the pairs of state posteriors of stereo features quite well to infer joint-state posteriors. This can be observed in the same gender and different gender cases with significant improvement. In the same talker condition, the explicit gain adjustment of the masker models provides a competitive result for the VTS based joint-decoder.

---

[2] - Available at https://github.com/skaae/DeepLearnToolbox

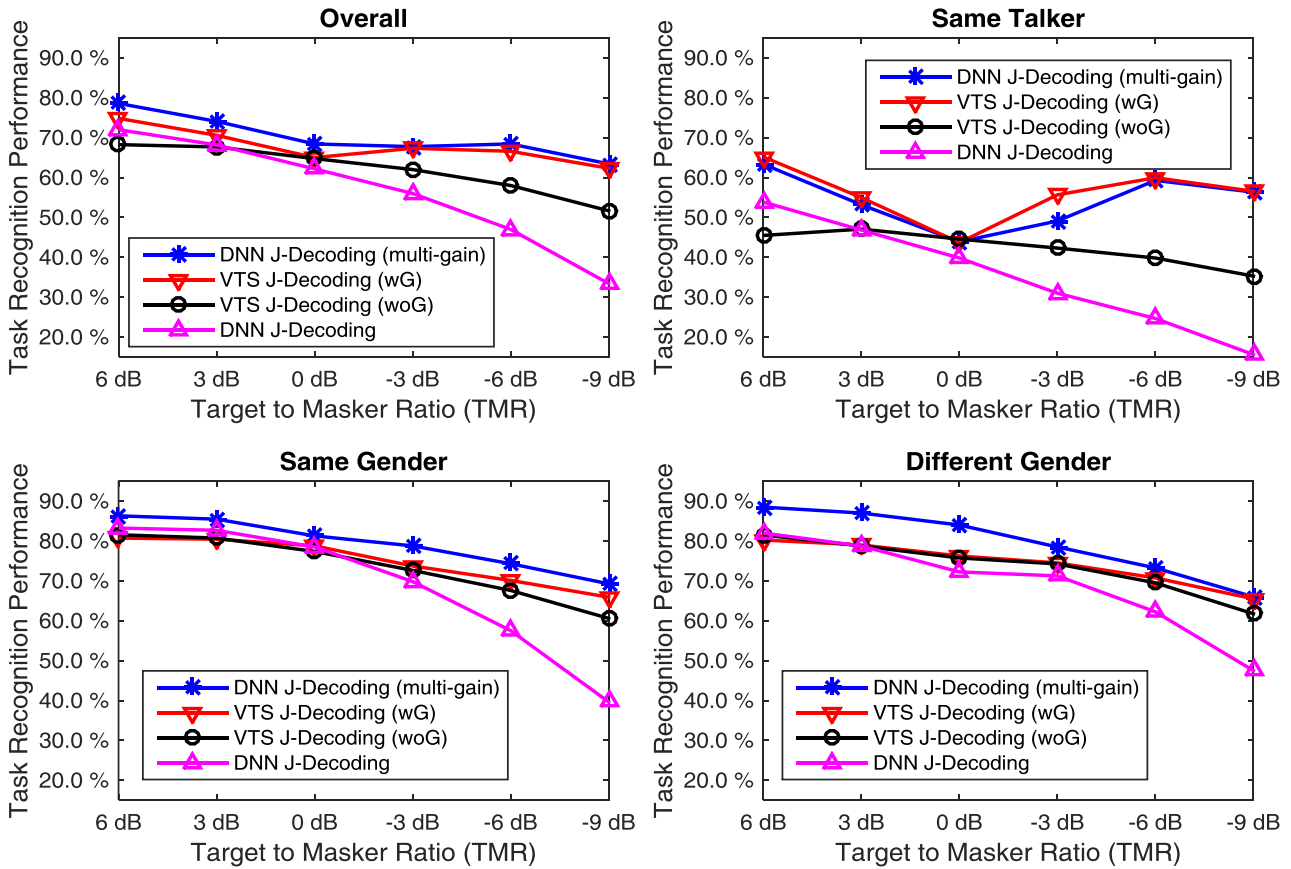

**Fig. 7.** Performance comparison of VTS and DNN based joint decoders. The comparison is done in different joint-speaker conditions: Same Talker, Same Gender, Different Gender, and Overall.

For the gain unaware cases in Fig. 7, we can see that the DNN based joint-posterior estimation is very sensitive to the mismatch between the training and test environments. Moreover, the DNN based joint-decoder performance degrades significantly in low TMR conditions when it cannot extract appropriate joint-state posteriors for decoding. As a final observation, we see that in the same talker condition, the dominant speaker marginal posterior information (extracted from the joint-state posteriors) helps improve decoding of the target speaker utterances even with misinformation extracted from the masker speaker dynamics.

Table 2 also lists Fig. 7's detailed results. In this table, we see that using information of pairs of state posteriors when training the DNNs improves the overall task results.

**Table 2: Performance comparison of VTS and DNN based joint-state decoders.**

| Test Case \ TMR (dB) | 6 | 3 | 0 | -3 | -6 | -9 | Overall |
|---|---|---|---|---|---|---|---|
| DNN based joint decoding (multi-gain trained DNN) | 78.6% | 74.1% | 68.4% | 67.8% | 68.4% | 63.4% | 70.1% |
| VTS based joint-decoding (adjusting masker gains) | 74.8% | 70.6% | 65.0% | 67.3% | 66.6% | 62.3% | 67.8% |
| VTS based joint-decoding | 68.3% | 67.7% | 64.8% | 62.0% | 58.0% | 51.7% | 62.1% |
| DNN based joint decoding | 72.0% | 68.2% | 62.2% | 56.0% | 47.0% | 33.4% | 56.5% |

For investigating the effectiveness of factorial speech processing models in the "monaural speech separation and recognition challenge" results of the IBM super human system [6], Microsoft research DNN-based [semi] joint-decoder [12] and LIMP VTS-based joint decoder [7] are provided in Fig. 8-left. These methods are factorial-based with different inference schemas. The IBM system first separates two speech sources using factorial models, then it performs single chain decoding; the Microsoft and LIMP systems directly perform joint decoding with two differences. First, the LIMP system is VTS-based, while the Microsoft system is DNN-based. Second, the LIMP system calculates joint-state likelihoods, while Microsoft extracts two separate marginal posteriors. This is why it is referred to as [semi]joint-decoder. As seen, all factorial based systems perform better than human listeners, and the IBM system won the challenge. Even the next best system [17] in the challenge was a factorial based system that approximately decodes utterances using a mega-state HMM decoder. All other results are far behind the super human systems [13].

To compare our proposed method to the closest method in the challenge, the Microsoft DNN-based system, we re-implemented two variants of their system called Hi+Low DNN/HMM and [Semi] Joint-Decoder [12] using a smaller acoustic model size. The performance of the original variants in their system and the re-implementation results are provided in Fig. 8-right compared to our proposed system results. As expected, the performance of the re-implemented variants is lower than the performance of original systems due to the acoustic models' simplicity in the re-implemented systems. Additionally, the results of the completely separate decoded utterances in Hi+Low DNN/HMM systems are surprising in extreme TMR conditions despite their poor results in near zero TMRs. For detailed configuration and discussion about the system variants, the reader can refer to [12]. The main observation from Fig. 8-right is that the joint-decoding performance using the proposed method for extraction of joint-posterior performs is better than the [semi] joint-decoder. The allowance of information passing between the two chains of factorial models in the true joint-decoding scenario gains more from the generative power of factorial models in conditions like the paper challenge. This cannot be exploited in separate posterior extraction as happened in the [semi] joint-decoder.

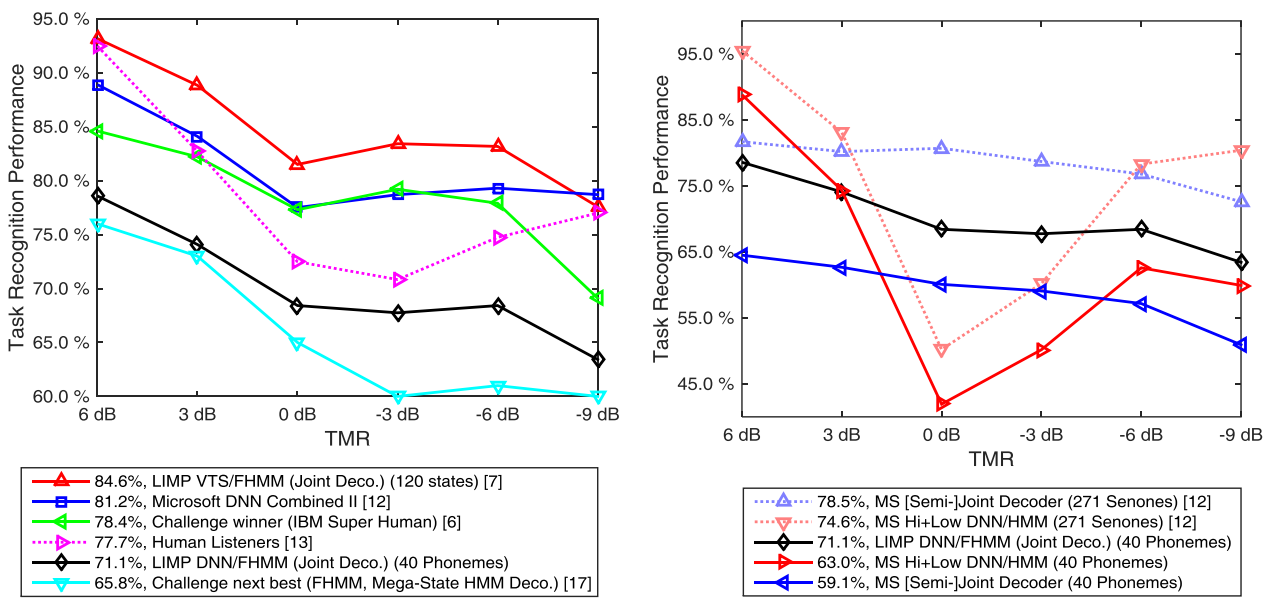

**Fig. 8. Left:** Performance comparison of the monaural speech separation and recognition challenge winner, other state of the art systems, and this paper on the challenge dataset.
**Right:** Investigating the effect of state space size in the performance of factorial models. As shown, the performance of the Microsoft (MS) systems with 40 states is significantly lower than the original systems. This justifies the performance gap between the current paper results and the best challenge systems. As another observation, using 40 states in a fully joint-state DNN based system works better than two separate High/Low DNNs.

# 5. Conclusion

An effective technique using deep neural networks was proposed for estimating joint-state posteriors of mixed-audio features to be used in factorial speech processing models. In the decoding phase of factorial model recognition applications, the decoder requires joint-state posteriors/likelihoods. We described four previous methods for calculating these posteriors where each method has its own drawbacks. The main tradeoff between the methods is the accuracy of joint-state posteriors and the computational complexity of joint-state posterior estimation. Our experiments show the proposed method's effectiveness compared to the only practical joint-state posterior estimation technique, the VTS method. The proposed method effectively exploits information of the pairs of state posteriors of stereo features to extract joint-state posteriors of mixed features for passing to the decoder.

This work's proposed method for training DNNs to extract joint-state posteriors is a way to construct DNN/FHMM speech recognizers that can also be used in noise-robust ASR applications. The ability to learn complex input-output mappings along with exploiting discriminating features of factorial chains makes the alternative to use two separate DNNs for extracting marginal posteriors a serious competitor to the proposed method for DNN/FHMM applications. The paper shows the superiority of joint posterior information in the joint-decoder compared to having two separate and even very accurate marginal posteriors. By using two separate DNNs for extracting marginal posteriors, we cannot exploit chain state dynamics information in the decoding which can be utilized by the proposed method.

The main developing aspect of the current work is to find a method for mitigating the quadratic increase in the number of joint-state neurons due to increases in source model states, which exponentially increases the network parameters.

## Acknowledgment

The authors would like to thank to Dr. M.A. Keyvanrad and S.S. Torabzadeh for their valuable arguments and suggestions in this work.